\documentclass[aps,twocolumn,prl,preprintnumbers,amsmath,amssymb,superscriptaddress,floats]{revtex4-1}

\usepackage{graphicx}
\usepackage{bm}
\usepackage{hyperref}
\usepackage[english]{babel}

\begin{document}

\title{Anomalous frequency and temperature dependent scattering and Hund's coupling in the almost quantum critical heavy fermion system CeFe$_2$Ge$_2$}

\author{G. Boss\'e}
\affiliation{The Institute for Quantum Matter, Department of Physics and Astronomy, The Johns Hopkins University, Baltimore, MD 21218 USA.}
\author{LiDong Pan}
\affiliation{The Institute for Quantum Matter, Department of Physics and Astronomy, The Johns Hopkins University, Baltimore, MD 21218 USA.}
\author{Yize S. Li}
\affiliation{Department of Physics and Fredrick Seitz Materials Research Laboratory, University of Illinois at Urbana-Champaign, Urbana, Illinois, 61801 USA.}
\author{L. H. Greene}
\affiliation{Department of Physics and Fredrick Seitz Materials Research Laboratory, University of Illinois at Urbana-Champaign, Urbana, Illinois, 61801 USA.}
\author{J. Eckstein}
\affiliation{Department of Physics and Fredrick Seitz Materials Research Laboratory, University of Illinois at Urbana-Champaign, Urbana, Illinois, 61801 USA.}
\author{N. P. Armitage}
\affiliation{The Institute for Quantum Matter, Department of Physics and Astronomy, The Johns Hopkins University, Baltimore, MD 21218 USA.}

\date{\today}

\begin{abstract}

We present THz range optical conductivity data of a thin film of the near quantum critical heavy fermion compound CeFe$_2$Ge$_2$.  Our complex conductivity measurements find a deviation from conventional Drude-like transport in a temperature range previously reported to exhibit unconventional behavior.  We calculate the frequency dependent effective mass and scattering rate using an extended Drude model analysis.  We find the inelastic scattering rate can be described by a temperature dependent power-law $\omega^{n(T)}$ where $n(T)$ approaches $\sim1.0 \pm 0.2$ at 1.5 K.   This is compared to the $\rho \sim T^{1.5}$ behavior claimed in dc resistivity data and the $\rho \sim T^{2}$ expected from Fermi-liquid theory.  In addition to a low temperature mass renormalization, we find an anomalous mass renormalization that persists to high temperature.  We attribute this to a Hund's coupling in the Fe states in a manner similar to that recently proposed in the ferro-pnictides.   CeFe$_2$Ge$_2$ appears to be a very interesting system where one may study the interplay between the usual $4f$ lattice Kondo effect and this Hund's enhanced Kondo effect in the $3d$ states.
\end{abstract}



\maketitle
\section{Introduction}
Many heavy fermion (HF) systems can be described within the framework of Fermi liquid (FL) theory, with quasiparticles exhibiting renormalized masses that are sometimes up to three orders of magnitude larger than bare electrons\cite{Basov11rev, Degiorgi99}. However, it is not universal that these systems always obey FL predictions; many HF systems deviate from FL response for electrical resistivity and specific heat\cite{Lohneysen94a,Schneider13}.  One explanation of such non-FL properties involves the proximity to magnetic order and a quantum critical point \cite{Andraka91, Kimura06, Mena05}.  Quantum criticality is seemingly ubiquitous across the diverse landscape of correlated matter and at the forefront of current research in novel phenomena of the cuprates, ruthenates and iron pnictides \cite{Lohneysen07a}.  Probing the emergent state of matter near a quantum critical point can, then, lead to further understanding of unsolved challenges, such as in high temperature superconductivity or hidden order phases.

There is a burgeoning list of HF materials that can be tuned toward quantum criticality. CeFe$_2$Ge$_2$ (CFG) is a HF compound with a non-magnetic ground state and a moderately enhanced heat capacity of 210 mJ/mol$\cdot$K$^2$ \cite{Ebihara1995219}.  It exhibits a metamagnetic anomaly at 300 kOe \cite{Sugawara1999} and is also believed to be close to a quantum critical point.  This behavior is reminiscent of other 4f-electron systems: CeRu$_2$Si$_2$ \cite{Haen1986}, CeNi$_2$Ge$_2$ \cite{Fukuhara1996} and CeCu$_6$ \cite{Schroder199247,Bosse12a}.  In addition to metamagnetic anomalies, deviations from ordinary FL behavior occur under certain circumstances \cite{Steglich1996, Aoki1998271, Bogenberger1995} in these materials.  CFG has been reported to show deviations from FL predictions; a T$^{1.5}$ dependence of resistivity in the temperature range $\sim$2-15 K is observed instead of the  FL T$^2$ dependence.  A crossover to a typical FL ground state was reported below a temperature T$_{FL} \approx2$ K \cite{Sugawara1999}.  By examining the Ce(Ru$_{1-x}$Fe$_x$)$_2$Ge$_2$ series, a phase diagram was constructed which shows that CFG is in proximity to a quantum critical point (QCP)\cite{Bud'ko1996111, Walf99a}.  In this series, ferromagnetic CeRu$_2$Ge$_2$ transitions to antiferromagnetic order around x=0.3 and then goes through a QCP at x=0.9 into a paramagnetic state.  The existence of a QCP  near the CFG end of the series could explain the non-FL behavior previously mentioned.

In this paper we use time domain terahertz spectroscopy (TDTS) to study the low-frequency complex conductivity of thin films of CFG and calculate the renormalized frequency dependent scattering rate and mass using an extended Drude model analysis.  Previously, the deviation from FL behavior in HF materials has been revealed primarily in the temperature dependence of the resistivity.  How this non-FL behavior may be seen in an ac technique is not as clear, but keeping within a quasi-Boltzmann transport point of view, a natural analog to the temperature dependence of resistivity may be the frequency dependence of the scattering rate. Therefore, deviations from FL $\omega ^2$ dependence can be analyzed by fitting the frequency dependent scattering rate using a power law constrained by the dc scattering rate.

\begin{figure}
\begin{center}
\includegraphics[width=1\columnwidth]{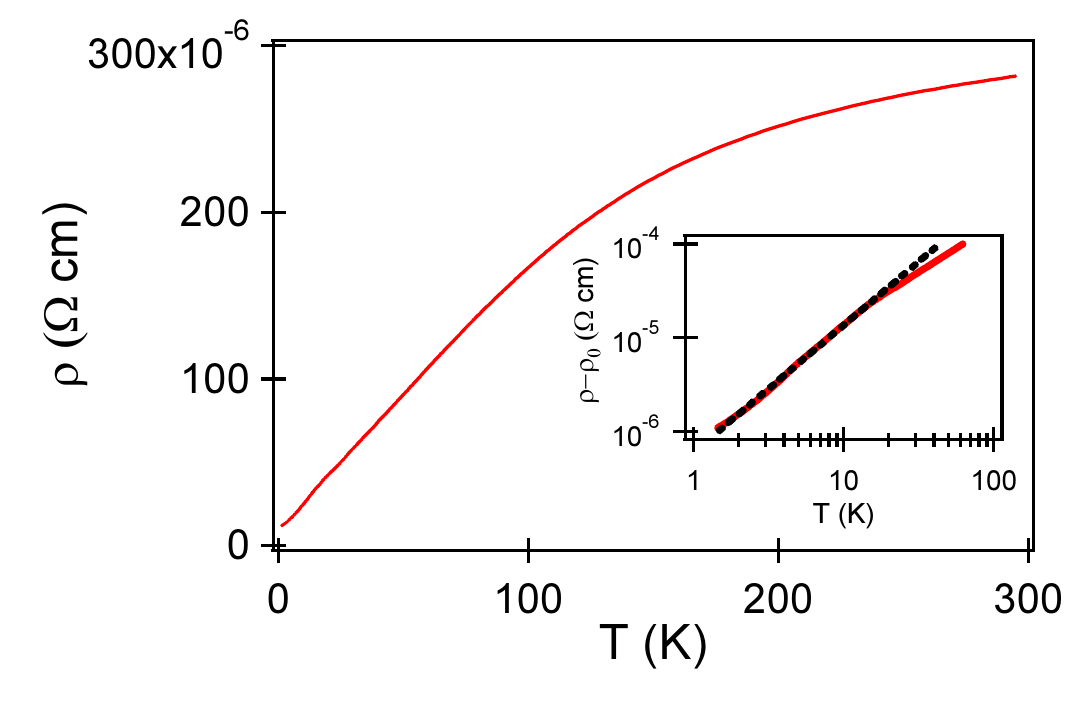}
\centering
\caption{dc resistivity as a function of temperature for the CeFe$_2$Ge$_2$ film.  Inset: Resistivity minus residual resistivity as a function of temperature.  A fit to the data in the temperature range 2-15 K is shown as a dashed line. }
\label{fig1}
\end{center}
\end{figure}

\section{Experimental Details}
In TDTS an infrared femtosecond laser pulse is split between two paths and sequentially excites a pair of  ``Auston" switch photoconductive antennas.  The first switch generates the THz pulse, which then travels through the sample. The second antenna receives the THz pulse and measures both the phase and amplitude of its electric field.  By dividing the Fourier transform of transmission through the sample by the Fourier transform of transmission through a reference substrate, the full complex transmission $T(\omega)$ as function of frequency can be obtained over a frequency range as broad as 100 GHz to 3.5 THz.  Details can be found elsewhere \cite{grischkowsky1990, Auston84, Lee09}.  The complex transmission is used to calculate the complex conductivity $\sigma(\omega)$ without the need for Kramers-Kronig transformation using the expression $T(\omega)= \frac{(1 + n)}{1+n + \sigma(\omega) d Z_0}\cdot e^{\frac{i\omega\Delta L(n-1)}{c}}$.  In this expression $n$ is the index of refraction of the substrate, $\Delta L$ is a correction factor that accounts for thickness  differences between the reference substrate and the sample substrate, $d$ is the film thickness, and $Z_0$ is the impedance of free space (377 $\Omega$).  The films studied in this work were grown by molecular beam epitaxy to a thickness of 524 $\AA$ using flux-matched codeposition on MgO substrates \cite{Yi2011a}.


In Fig. \ref{fig1} we show dc resistivity which decreases monotonically as temperature is lowered.  Around 15K there is a subtle crossover that can be better seen in the inset of Fig. \ref{fig1}, where the zero temperature residual resistivity $\rho(0)$ has been subtracted off. We find $\rho(T)-\rho(0)$ $\propto$ $T^{1.4}$ for temperatures between $\sim$2-15K, which is consistent with previous measurements \cite{Yi2011a}.  A $T^{1.5}$ dependence of the resistivity was previously explained as an anomalous property around a magnetic instability, similar to other HF systems\cite{Moriya95a}.

\begin{figure}
\begin{center}
\includegraphics[width=1\columnwidth]{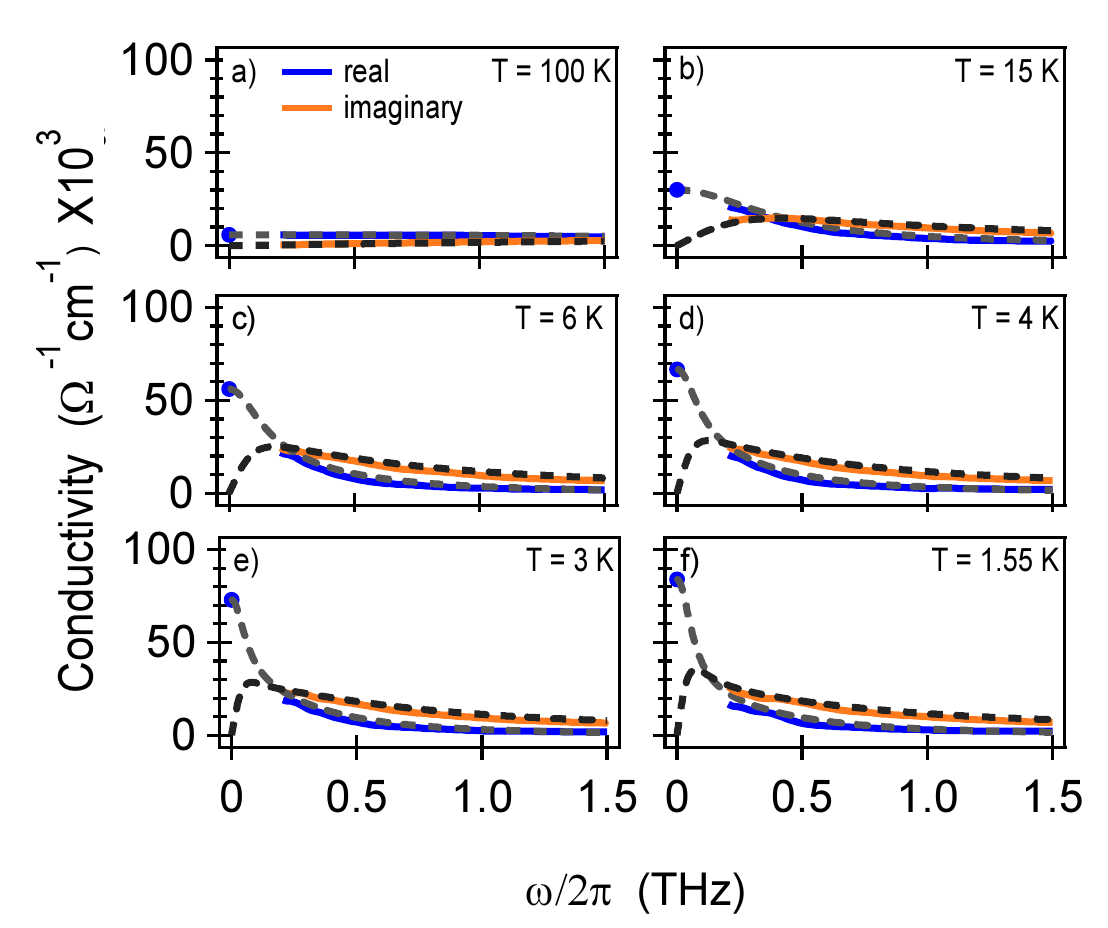}
\centering
\caption{(Color online) (a-f) Real and imaginary parts of the complex conductivity as functions of frequency for select temperatures.  Solid lines indicate experimental data, while dashed lines show results of a two component Drude fit described in the text.  The values of the dc conductivity are shown as solid circles.}
\label{fig2}
\end{center}
\end{figure}

In Fig. \ref{fig2} we show the real and imaginary parts of the complex conductivity $\sigma(\omega)$ for a few temperatures (see Ref.\cite{BosseSM} for all data). At high temperatures both parts of the complex conductivity are flat and featureless.  As the temperature is lowered a shift in spectral weight toward low frequencies is seen in the real part of the conductivity $\sigma_1$.  A narrow Drude-like peak grows as the temperature is decreased to 1.55 K.  Using the dc data, the low-temperature terahertz (THz) conductivity can be fit to a Drude model using two zero-frequency oscillators and a high-frequency dielectric constant $\epsilon_\infty$, which accounts for high-frequency contributions above the measured spectral range \cite{Kuzmenko05a}.  This two Drude component fit should be interpreted as a simple Kramers-Kronig consistent parameterization of the data.   It does not necessarily correspond to two distinct charge carrier species\cite{BosseSM}.  Fits to the complex conductivity are shown as dashed lines in Fig. \ref{fig2}. These fits are highly constrained by the use of both THz and dc data.  Implicit in this fitting is the reasonable assumption that there are no spectral peaks in $\sigma_1$ below the lower end of our measured range \cite{Dressel02a,Scheffler05a,Bosse12a}.

\begin{figure*}
\begin{center}
\includegraphics[width=1\textwidth]{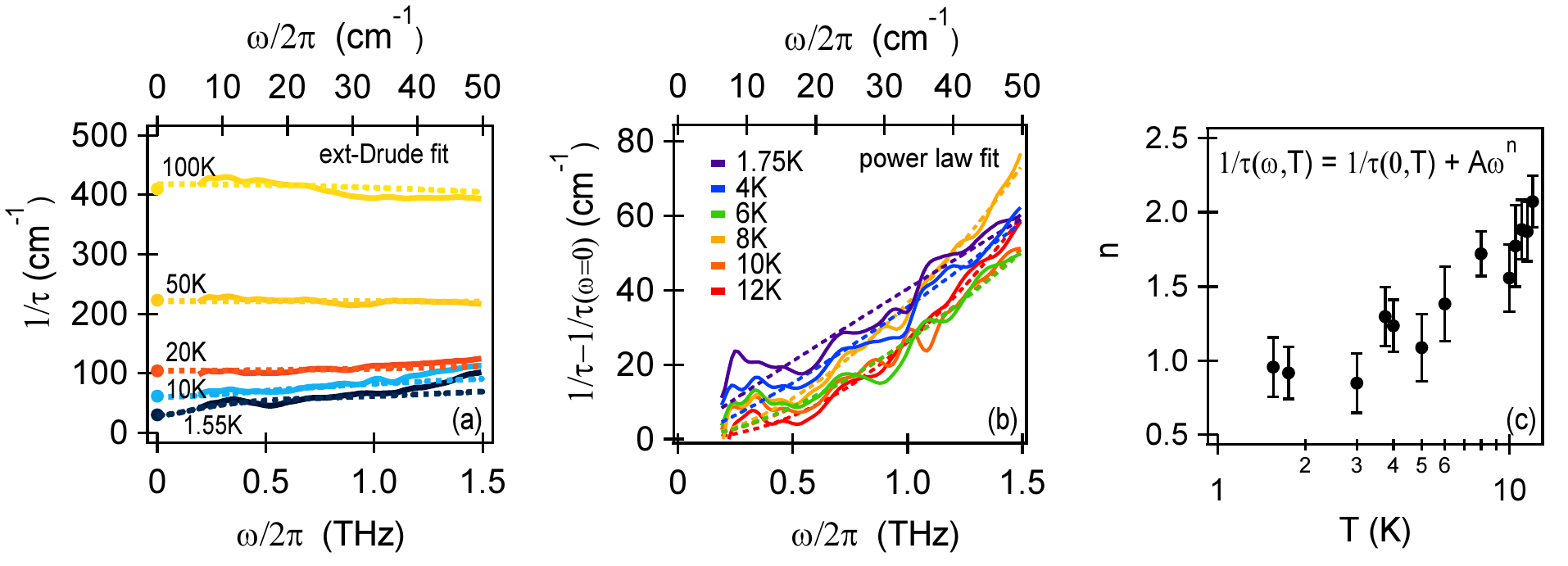}
\centering
\caption{(Color online) (a) Scattering rates as functions of frequency.  Experimental data are shown as solid lines; extended Drude model fits, described in the text, to the data are shown as dashed lines.  dc data points are shown as solid circles.  (b) Scattering rates minus dc values versus frequency with power law fits, according to the equation $\frac{1}{\tau(\omega,T)}= \frac{1}{\tau(0,T)} + A \omega^{n}$ for temperatures below 12 K.  (c) The power law exponents of the fits to the scattering rates as a function of temperature.  Error bars were estimated by additionally fitting the data from 0.2 THz to both 1.25 THz and 1.75 THz and taking the average of the values \cite{BosseSM}.}
\label{fig:fig3}
\end{center}
\end{figure*}


\section{Discussion}
The fact that we cannot fit our data with a single term Drude model demonstrates that the transport cannot be described by a single energy-independent relaxation time.  An alternative to fitting the data with multiple Drude-like terms is the extended Drude model \cite{Allen77a}.  In this formalism, a frequency-dependent mass $m^*(\omega)/m_b$ and scattering rate $1/\tau(\omega)$ are extracted from the measured optical constants by the relations

\begin{eqnarray}
\label{mass}
\frac{m^*(\omega)}{m_{b}}=-\frac{\omega_{p}^2}{4\pi\omega}Im \left[\frac{1}{\sigma(\omega)} \right] \\
\frac{1}{\tau(\omega)}=\frac{\omega_{p}^2}{4\pi}Re \left[\frac{1}{\sigma(\omega)}\right] \label{scatrate}
\end{eqnarray}

\noindent where $\omega_{p}$ is the plasma frequency that is a measure of the total Drude spectral weight (proportional to $\omega_{p}^2$) of all free charge carriers and $m_b$ is the band mass.  To determine the total spectral weight of all the free charge carriers one must measure to much higher frequencies than the THz regime.  In this regard, we also performed Fourier transform infrared spectroscopy transmission and reflection measurements from 100 to 8000 cm$^{-1}$ from room temperature to 5 K.   With the TDTS results, we fit the full measured spectral range (from 0 to 8000 cm$^{-1}$) to a generalized Kramers-Kronig consistent Drude/Drude-Lorentz model\cite{BosseSM}.  As discussed in the Supplementary Information, we partitioned the spectral weight into low frequency ``intraband" (e.g. Drude) carriers and higher frequency ``interband" excitations\cite{BosseSM}.  The plasma frequency of the intraband piece is $\omega_p$ = 2$\pi \times (12150 \pm 1000)$ cm$^{-1}$ = 2$\pi \times (365 \pm 30)$ THz.  The uncertainty on the absolute numerical value of $\omega_{p}^2$ arises from uncertainties in assigning spectral weight to either free carrier Drude or to finite frequency Drude-Lorentz oscillators. While this uncertainty sets the overall scale of the renormalized optical quantities, it does not affect their frequency or temperature dependence.

In Fig. \ref{fig:fig3}(a) we show the frequency dependence of the scattering rate for a few temperatures between 1.55-100 K calculated using Eq. \ref{scatrate}.  The dc points are calculated using the plasma frequency obtained above and resistivity data.  Fits to the data, shown as dashed lines, were created by inverting the 2 Drude component model parameterizations of the conductivity, using extended Drude model equations\cite{BosseSM}.  These fits serve as a Kramers-Kronig compatible way to connect the THz data to the dc data.   The detailed functional dependence should not be viewed as exact, but the general dependence is valid.  For temperatures below 15K, on top of a large constant offset, a frequency dependence develops to the scattering rate.  We fit the scattering rate data to a power law $\frac{1}{\tau(\omega,T)}= \frac{1}{\tau(0,T)} + A \omega^{n}$  where $\frac{1}{\tau(0,T)}$ is the dc scattering rate and $n$ is a temperature dependent exponent.  The scattering rate minus the dc value is plotted versus frequency with a corresponding power law fit for a few temperatures below 12 K in Fig. \ref{fig:fig3} (b). These fits are constrained to fit data from 0.2-1.5 THz and pass through zero at zero frequency.  In Fig. \ref{fig:fig3}(c) we show the power law exponent of the fits shown in Fig. \ref{fig:fig3}(b).

In the temperature range previously known to exhibit non-FL behavior, we observe patterns that suggest a deviation from conventional FL dependence: for temperatures below $\sim$3 K the power law exponents are on average 0.94 $\pm$ 0.02.  From 3-10 K the power law exponent averages 1.30 $\pm$ 0.27. Above 10 K the exponent is close to 2.  Errors in the power law fits mostly originate in determining over which frequencies to fit and, unfortunately, make it difficult to distinguish whether there is a unequivocal difference between the frequency dependence found here and the $\sim T^{1.5}$ behavior reported for the dc resistivity  \cite{Sugawara1999} at temperatures between 1.5 and 10 K.  The presence of different exponents appearing for the temperature and frequency dependencies of the scattering rate is anomalous and does not find ready explanation in the Hertz-Millis-Moriya (HMM) model of quantum criticality above the critical dimension \cite{Hertz76a, Millis93a, Moriya95a, Lohneysen07a,Rosch99a}.  Despite the seeming conflict, we note that this difference in the functional dependence of the optical self-energy is similar to the differences in the functional dependence expected within the HMM formalism for the susceptibility itself  \cite{Schroder00a, Singh11a}. In view of this, a possible connection between these quantities is left open for future theoretical models.   An alternative explanation is that the T$^{1.5}$ dependence of the dc resistivity seen in our data and previous reports, in fact, reflects a crossover from a larger exponent to the approximately 1 we find at 1.55K.   With this interpretation, we would find the same approximately linear dependence for T and $\omega$ in both low energy limits.  This may be consistent with some models for HMM criticality in clean systems \cite{Rosch99a}.  Whichever the case, our data provides further evidence that there is a deviation from canonical FL theory which predicts the scattering rate's dependence on frequency and temperature to differ only by a factor of 4$\pi^2$ via the relation $\frac{1}{\tau(\omega, T)}= A[(\hbar\omega)^2+ (2\pi k_BT)^2]$ \cite{Berthod13}.  While our data could be described by a somewhat modified version of this equation, in which the coefficient of the temperature term differs from 4$\pi^2$, with quadratic dependencies it is only possible to fit the data in a very narrow frequency range, reinforcing the idea that FL theory is not a good description at this energy scale.

\begin{figure}
\begin{center}
\includegraphics[width=1\columnwidth]{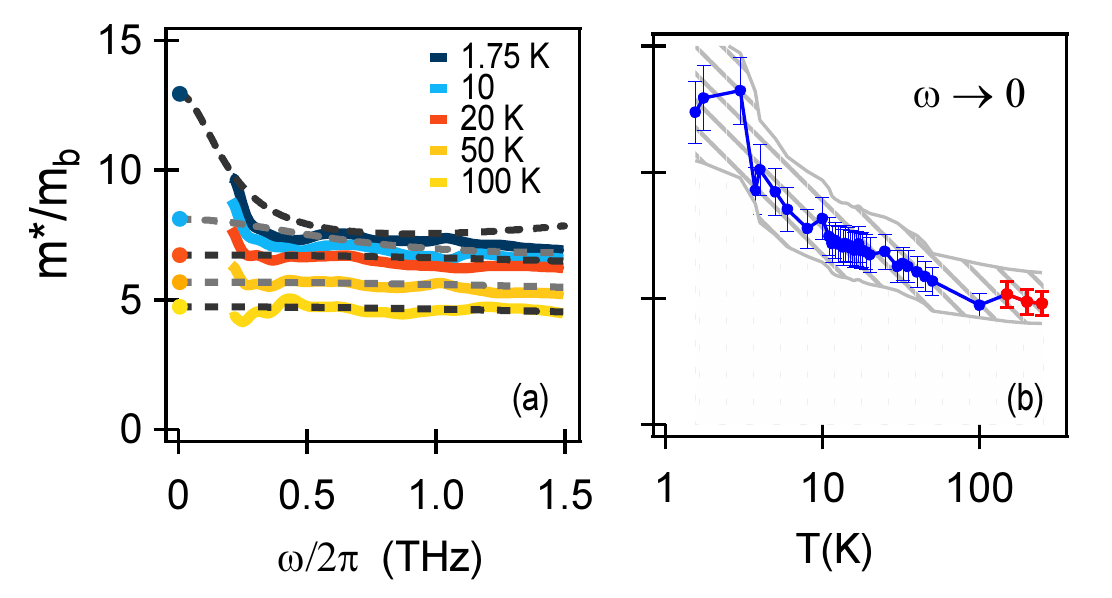}
\centering
\caption{(Color online) (a)Renormalized masses as functions of frequency. (b) The $\omega\rightarrow 0$ limit of the frequency dependent renormalized masses as functions of temperature extracted from extended Drude model fits. Error bars reflect uncertainty in the extended Drude model fits, while the grey shadow reflects the uncertainty from the range of possible values of $\omega_p$.  The red points are from a dataset that was taken at an earlier time than the data represented in blue.}
\label{fig:fig4}
\end{center}
\end{figure}

Continuing with our extended Drude model analysis, we show the renormalized mass as a function of frequency and the $\omega\rightarrow 0$ limit of the masses as a function of temperature in Fig. \ref{fig:fig4}.  The renormalized mass does not exhibit strong dependence in the measured frequency range; the renormalization is largest below 0.3 THz, which is just above the cut-off of our data.  However, because we know the dc scattering rate and the renormalized mass is related to this via Kramers-Kronig, we can determine how the mass interpolates to zero frequency using an extended Drude model fitting technique which simultaneously fits the scattering rate and mass\cite{BosseSM}.

We would like to note that one powerful feature of the TDTS technique is its ability to measure the temperature dependence of the mass renormalization. Other techniques (e.g. heat capacity and quantum oscillations) that have contributed to the study of HFs typically only extract the mass renormalization at the lowest temperatures through the temperature dependence itself once the coherent state has developed.   In contrast, TDTS allows us to see the formation of the heavy state; with the extended Drude model analysis we can measure the mass renormalization from high temperatures all the way into the coherent state.  In this regard, one particularly notable feature of this data is the mass renormalization that persists as high as $m^*/m_b \sim 5$ at high temperature, as seen in Fig. \ref{fig:fig4}(b).   The overall scale of the mass is set by the measured plasma frequency and although it is possible that we have parameterized it inaccurately, the present value of the low temperature renormalized mass is consistent with heat capacity measurements and materials with similar specific heat coefficients\cite{Ebihara1995219,Yi2011a}.

 We believe that this high temperature mass enhancement follows from a ``Hund's coupling enhanced" Kondo effect as proposed for ferropnictide compounds with related crystal structures and composition \cite{Nevidomskyy09a,Haule09a,Johannes09a,Medici11a}. In the case of KFe$_2$As$_2$, local density approximation + dynamical mean field theory calculations suggest that the mass enhancement in iron pnictides occurs because the electrons, which are somewhat localized at high temperature due to Hund's rule coupling, form coherent quasiparticle bands with the underlying Fermi surface\cite{Hardy13}.  Further Gutzwiller corrected electronic structure calculations for this material support the idea of an ``orbital selective" Mott transition, in which localization due to strong Coulomb repulsion occurs for only some of the orbitals\cite{Hardy13}.  Qualitative aspects of these calculations should apply to the present case, in which Fe$^{+2}$ is found in the same 122 crystal structure.  The interplay of the localized and delocalized bands can then give rise to behavior very similar to that of 4$f$-based heavy fermion systems, even though all states formed in KFe$_2$As$_2$ are from Fe 3$d$ orbitals\cite{Hardy13}.  With that in mind, it is likely that a strong Hund's interaction in the $3d$ Fe atoms creates a mass enhancement at temperatures higher than naively expected from the usual treatment of the $4f$ Ce moments hybridizing with the conduction band.  Bolstering this interpretation is the observation that both the $4f$ moment free compounds of LaFe$_2$Ge$_2$ and LaFe$_2$Si$_2$ show anomalously large low temperature specific heat coefficients of 37 mJ/mol$\cdot$K$^2$ and 22.7 mJ/mol$\cdot$K$^2$ respectively \cite{Ebihara1995219,Svoboda03a}.  These can be compared to La 122 compounds based on wider d-band materials  with Ru and Cu that have more conventional magnitudes of 5-7 mJ/mol$\cdot$K$^2$ \cite{Besnus85a,Gondek11a}.  CFG (with a heat capacity of 210 mJ/mol$\cdot$K$^2$) then appears to be a very appealing system to study the interplay between the usual $4f$ Kondo effect and this Hund's enhanced Kondo effect in the $3d$ states.

\section{Conclusion}
In summary, we have applied time domain THz spectroscopy to the heavy fermion compound CeFe$_2$Ge$_2$ and investigated the non-FL behavior of its optical properties using an extended Drude model analysis.  Evidence was found in the frequency dependent scattering rate to support previous suggestions that CFG shows non-FL behavior in the temperature range 2-15K.   We find a crossover in the inelastic scattering rates that can be described by a power-law $\omega^{n}$ where $n$ becomes approximately unity in the same temperature range where the dc resistivity is reported to show  $T^{1.5}$ dependence.   Counter to the usual Kondo scheme for Ce based compounds, we find that the mass enhancement persists as high as 250K, an effect that we believe originates in a ``Hund's coupling enhanced" Kondo effect.  CeFe$_2$Ge$_2$ seems to be an interesting system where one may investigate the interplay between the standard $4f$ lattice Kondo effect and this Hund's enhanced Kondo effect in the $3d$ states.    In further experiments it would be interesting to compare LaFe$_2$Ge$_2$  with LaCu$_2$Ge$_2$  and LaRu$_2$Ge$_2$ to isolate the effect of Hund's coupling.

We would like to thank P. Coleman, G. Kotliar, J. Paglione, F. Ronning, and A. Rosch  for helpful conversations.   THz conductivity measurements at Johns Hopkins University were supported by the Gordon and Betty Moore Foundation through Grant No. GBMF2628 to NPA.   The film growth at the University of Illinois at Urbana-Champagne was supported by the Center for Emergent Superconductivity, an Energy Frontier Research Center funded by the U.S. Department of Energy, Office of Science, Office of Basic Energy Sciences under Award No. DE-AC0298CH1088.

\bibliography{cfgreferences_final}

\end{document}